\documentclass[aps,onecolumn,pra,longbibliography]{revtex4}
\usepackage{graphicx,amsmath,color}
\usepackage{float}
\usepackage[dvipdfmx, colorlinks=true, linkcolor=blue, urlcolor=blue, citecolor=blue]{hyperref}
\usepackage{bm}
\usepackage{amsfonts}
\usepackage{here}

\begin{document}

	\title{Orthogonal and antiparallel vortex tubes and energy cascades in quantum turbulence}
	\author{Tsuyoshi Kadokura}
	\affiliation{
		Department of Engineering Science, University of Electro-Communications, Tokyo 182-8585, Japan
	}
	\author{Hiroki Saito}
	\affiliation{
		Department of Engineering Science, University of Electro-Communications, Tokyo 182-8585, Japan
	}
	\date{\today}
	\begin{abstract}
		We investigate the dynamics of energy cascades in quantum turbulence
		by directly observing the vorticity distributions
		in numerical simulations of the Gross-Pitaevskii equation.
		By Fourier filtering each scale of the vorticity distribution,
		we find that antiparallel vortex tubes at a large scale generate
		small-scale vortex tubes orthogonal to those at the large scale,
		which is a manifestation of the energy cascade from large to small scales.
		We reveal the dynamics of quantized vortex lines in these processes.
	\end{abstract}
	%
	%
	\maketitle

	\section{INTRODUCTION}
	\label{s:introduction}
	As the Reynolds number increases, laminar fluid flow develops into turbulence
	due to hydrodynamic instability~\cite{O.Reynolds},
	and energy is transferred from large to small scales.
	Such an energy cascade has already been implied in the famous sketch of
	water turbulence drawn by Leonardo da Vinci,
	in which many large and small vortices
	are tangled with each other.
	More specifically, the energy cascade in turbulence
	was illustrated by Richardson, in which large-scale vortex rings
	are divided into
	small-scale rings~\cite{L.F.Richardson}.
	By a statistical approach with universality assumptions,
	the energy cascade has been shown to lead to Kolmogorov's $-5/3$ power law
	in the energy spectrum, which has been thoroughly investigated theoretically
	and observed experimentally~\cite{U.Frisch1, A.N.Kolmogorov, G.K.Batchelor,
	T.Tatsumi, R.H.Kraichinan, U.Frisch2, S.Goto1, S.Goto2}.

	Quantum fluids are different from classical fluids in that,
	in quantum fluids, vortices are quantized and viscosity is absent.
	Despite these differences, it has been shown that classical
	and quantum fluids share a variety of
	hydrodynamic phenomena~\cite{K.Sasaki1, M.T.Reeves1, W.J.Kwon, K.Sasaki2,
	T.Kadokura, H.Takeuchi, A.Bezett}.
	Energy cascades and power-law spectra have been observed
	in quantum turbulence of superfluid helium~\cite{L.Skrbek}
	and ultracold atomic gases~\cite{N.Navon}.
	Theoretically, it has been revealed that turbulent quantum fluids
	exhibit Kolmogorov's power law in incompressible kinetic-energy
	spectra~\cite{M.Tsubota1, M.Kobayashi1, M.Kobayashi2, M.Kobayashi3}.
	A variety of dynamics and power-law spectra in quantum fluids have been
	investigated~\cite{A.W.Baggaley, M.Tsubota2,
	K.Fujimoto, P.J.Tabeling, A.C.White, A.Villois, P.M.Walmsley, M.T.Reeves2,
	C.F.Barenghi1, C.F.Barenghi2, M.Kursa, R.M.Kerr1, S.R.Stalp, K.W.Madison,
	N.G.Parker, C.Nore}.

	Recently, Goto {\it et al}.~\cite{S.Goto1, S.Goto2}
	numerically studied the energy cascade in classical fluids
	from the perspective of the vorticity distribution.
	By taking the Fourier transform of the vorticity distribution and
	applying a band-pass filter to
	the Fourier components, they extracted each scale of the vorticity
	distribution of the turbulent flow. They found that the vortex tubes
	at each scale tend to align in antiparallel, while vortex tubes
	generated at smaller scales tend to be orthogonal to the antiparallel
	vortex tubes at larger scales. This dynamics leads to the energy
	transfer from large to small scales, and may be responsible for
	the energy cascade and Kolmogorov's power law.
	Since there are various similarities between
	the hydrodynamic phenomena in classical and quantum fluids, we expect that
	the dynamics observed at each hierarchy in the vorticity distribution
	of classical fluids should also be observed in quantum fluids.
	We investigate this correspondence in the present study.

	In the present paper, we focus on each level of hierarchy in the vorticity
	distribution in a quantum fluid.
	First, so that the expected dynamics can be clearly seen, antiparallel vortex bundles
	are imprinted artificially in a uniform superfluid.
	Using a method of Fourier filtering, these bundles of quantized vortices
	are visualized as vortex tubes. We find that
	antiparallel vortex tubes generate small-scale vortex tubes
	orthogonal to the antiparallel vortex tubes.
	Next, we apply our Fourier filtering scheme to isotropic fully-developed
	turbulence that exhibits Kolmogorov's $-5/3$ power law.
	By calculating the angles between the directions of the vortex tubes,
	we confirm the existence of antiparallel correlation at each scale
	and orthogonal correlations
	between different scales in the quantum turbulence,
	as observed for classical turbulence.

	This paper is organized as follows.
	Section~\ref{s:formulation} formulates the problem and describes the numerical method.
	The dynamics of the initially located vortex bundles are studied in Sec.~\ref{s:nucleation}.
	Section~\ref{s:distributions} studies the case of fully developed quantum turbulence.
	Conclusions are given in Sec.~\ref{s:conclusions}.

	\section{FORMULATION OF THE PROBLEM}
	\label{s:formulation}
	We consider the dynamics of a quantum fluid,
	which is described by the three-dimensional
	Gross-Pitaevskii equation given by
	\begin{eqnarray}
		\label{eq:GPE}
		(i-\gamma) \hbar \frac{\partial \Psi}{\partial t}
		= - \frac{\hbar^2}{2m} \bm{\nabla}^2 \Psi
		+ U({\bm r}, t)\Psi
		+ \mathrm{g} | \Psi |^2 \Psi,
	\end{eqnarray}
	where $\Psi({\bm r},t)$ is the macroscopic wave function,
	$m$ is the particle mass,
	$U({\bm r},t)$ represents an external potential,
	and $\mathrm{g}$ is the interaction coefficient.
	Equation~(\ref{eq:GPE}) includes a phenomenological dissipation
	constant $\gamma$~\cite{M.Kobayashi1,M.Kobayashi2, M.Kobayashi3}
	to eliminate large-wavenumber components
	generated by the energy cascade. We take $\gamma = 0.004$ in this study.
	We normalize the wave function as $\tilde{\psi}=n_0^{-1/2}\Psi$,
	where $n_0$ is the atomic density $|\Psi|^2$
	in a uniform system without $U$.
	The length and time are normalized
	by $\xi = \hbar/(m \mathrm{g} n_0)^{1/2}$ and $\tau = \hbar/({\rm g}n_0)$
	respectively.
	Equation~(\ref{eq:GPE}) is then normalized as
	\begin{eqnarray}
		\label{eq:NGPE}
		(i-\gamma) \frac{\partial \tilde{\psi}}{\partial \tilde{t}}
		= \left[
			- \frac{1}{2} \tilde{\nabla}^2
			+ \tilde{U}
			+ |\tilde{\psi} |^2
		\right] \tilde{\psi},
	\end{eqnarray}
	where $\tilde{\nabla}^2 = \xi^2 \bm{\nabla}^2$ and $\tilde{U} = U/({\rm g}n_0)$.

	The flux of the mass current is given by
	\begin{eqnarray}
		\label{eq:CURRENT}
		{\bm J} = \frac{1}{2 i}
		\left(
			\tilde{\psi}^\dagger \tilde{\nabla} \tilde{\psi} - \tilde{\psi} \tilde{\nabla} \tilde{\psi}^\dagger
		\right) = |\tilde{\psi}|^2 \tilde{\nabla} \phi,
	\end{eqnarray}
	where $\phi$ is the phase of $\tilde{\psi}$.
	The vorticity $\bm{\Omega}$ is usually defined as the curl
	of the velocity field,
	$\bm{\Omega} = \tilde{\nabla} \times (\bm{J} / |\tilde{\psi}|^2) = \tilde{\nabla} \times \tilde{\nabla} \phi$.
	The vorticity $\bm{\Omega}$ therefore vanishes everywhere
	except at the singularity of the quantized vortex core.
	To avoid the singularity in the vorticity distribution,
	we define the vorticity distribution of the mass current as
	\begin{eqnarray}
		\label{eq:VORTICITY}
		{\bm W} = \tilde{\nabla} \times {\bm J},
	\end{eqnarray}
	which is a smooth function even at the vortex core and
	numerically tractable.
	We extract a specific scale of the vorticity distribution by applying
	the band-pass Fourier filter~\cite{S.Goto2, S.Goto3},
	\begin{eqnarray}
		\label{eq:FFT}
		\tilde{{\bm W}}(\bm{k}, t; k_c) = \left\{
			\begin{array}{ll}
				\int {\bm W}({\bm r}, t) e^{-i \bm{k} \cdot {\bm r}}  {\rm d}\bm{r}
				& (k_c/\sqrt{2}< |\bm{k}| <\sqrt{2}k_c)
				\\
				\\
				0 & ({\rm otherwise})
			\end{array}
		\right.,
	\end{eqnarray}
	where $k_c$ is the characteristic wavenumber of the band-pass filter.
	The band-pass filtered vorticity distribution is thus given by
	\begin{eqnarray}
		\label{eq:IFFT}
		{\bm W}(\bm{r}, t; k_c) =
		\frac{1}{V} \sum_{\bm{k}} \tilde{\bm{W}}(\bm{k}, t) e^{i \bm{k} \cdot \bm{r}},
	\end{eqnarray}
	where $V=L^3$ is the volume of the system.
	Let us consider vorticity distributions $\bm{W}(\bm{r}, t; k_c)$ at two scales,
	$\bm{W}_1$ and $\bm{W}_2$,
	where $k_c$ for $\bm{W}_2$ is larger than that for $\bm{W}_1$.
	We define an angle $\theta_{12}$ between
	$\bm{W}_1(\bm{r})$ and $\bm{W}_2(\bm{r}+\Delta\bm{r})$ as
	\begin{eqnarray}
		\label{eq:ORTH}
		\cos \theta_{12}
		= \frac{{\bm W}_1({\bm r}) \cdot {\bm W}_2({\bm r} + \Delta {\bm r}) }
		{
			|{\bm W}_1({\bm r})| |{\bm W}_2({\bm r} + \Delta {\bm r})|
		}.
	\end{eqnarray}
	We also define an angle $\theta_{11}$ between
	$\bm{W}_1(\bm{r})$ and $\bm{W}_1(\bm{r}+\Delta\bm{r})$ as
	\begin{eqnarray}
		\label{eq:ANTI}
		\cos \theta_{11}
		= \frac{{\bm W}_1({\bm r}) \cdot {\bm W}_1({\bm r} + \Delta {\bm r}) }
		{
			|{\bm W}_1({\bm r})| |{\bm W}_1({\bm r} + \Delta {\bm r})|
		}.
	\end{eqnarray}
	In numerical calculations, we collect the values of
	$\cos \theta_{12}$ and $\cos \theta_{11}$
	by varying
	$\bm{r}$ and $\Delta\bm{r}$ where $|\Delta\bm{r}|$ is
	restricted to some range.
	We thus obtain the occurrence distributions
	$P_{12}(\cos \theta_{12})$ and $P_{11}(\cos \theta_{11})$ for these angles.
	These distributions thus correspond to the probability distributions
	of the angle between the vorticities, if we choose $\bm{r}$
	(and $\Delta \bm{r}$ within some range) randomly.
	If the vorticities $\bm{W}_1$ at $\bm{r}$ and $\bm{r} + \Delta \bm{r}$
	tend to be antiparallel with each other, $P_{11}(\cos \theta_{11})$
	has a peak at $\cos \theta_{11} = -1$.
	If the vorticity $\bm{W}_1(\bm{r})$ at a larger scale
	generates the vorticity $\bm{W}_2(\bm{r}+\Delta \bm{r})$ at a smaller scale
	and the latter tends to be orthogonal to the former,
	$P_{12}(\cos \theta_{12})$ has a peak at $\cos \theta_{12} = 0$.
	These tendencies in $\bm{W}_1$ and $\bm{W}_2$ will be
	numerically shown in the next section.

	We numerically solve Eq.~(\ref{eq:NGPE}) using the pseudospectral method,
	and therefore, a periodic boundary condition is imposed.
	The numerical space is taken to be $L^3=128^3$ with mesh of
	$\Delta x = \Delta y = \Delta z = 1$.
	\section{NUMERICAL RESULTS}
	\label{s:results}
	\subsection{Dynamics of vortex bundles}
	\label{s:nucleation}
	\begin{figure}[t]
		\centering
		\includegraphics[width=18cm]{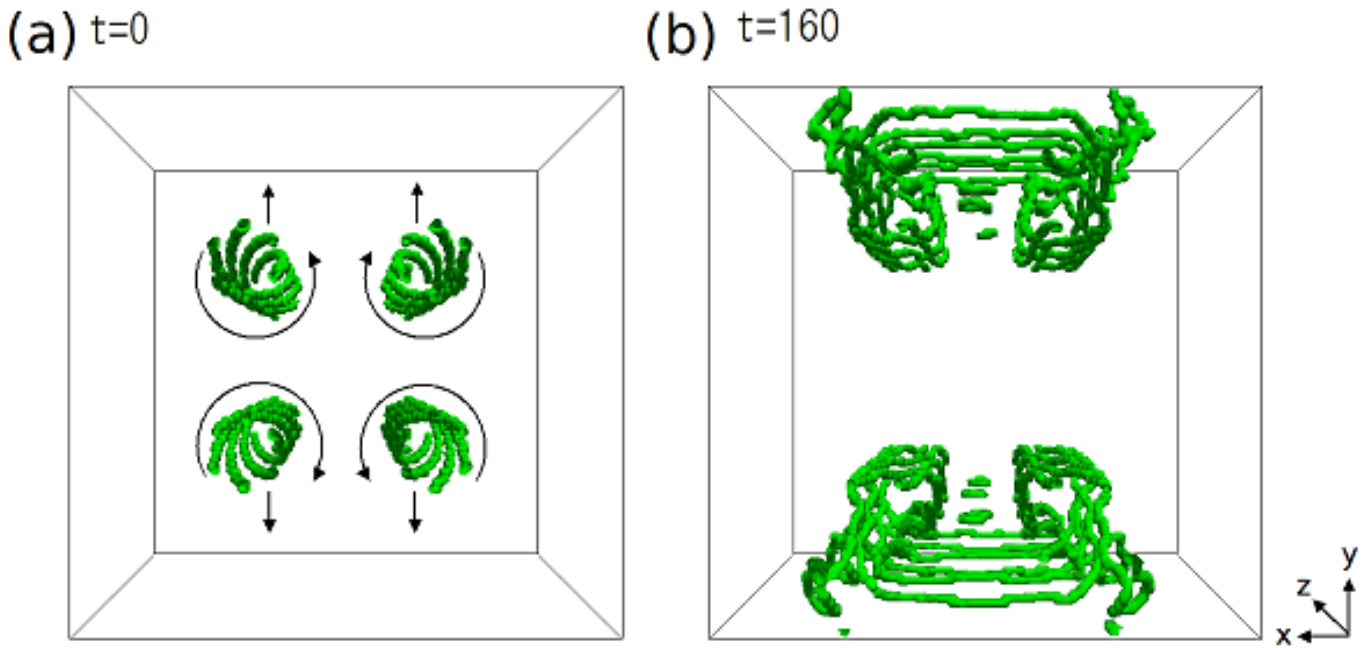}
		\caption{
			Dynamics of quantized vortex cores for an initial state
			with four vortex bundles. The numerical pixels around which
			the phase rotates by $2\pi$ are visualized.
			\
			(a) Four bundles with six vortex lines in each bundle
			prepared in a uniform system using Eqs.~(\ref{eq:SETUPEQ1})
			and~(\ref{eq:SETUPEQ2}) with $r_b=16$. The centers of
			the bundles are located at
			$(x,y)=(64,64),(108,64),(64,108),$ and $(108,108)$.
			\
			The curly arrows indicate the rotation directions of the vortices
			and the arrows in the $\pm y$ directions indicate the directions of
			their propagation.
			(b) Snapshot at $t=160$. Two pairs of vortex bundles travel in the $\pm y$ directions,
			and vortex rings and ladder structures are formed.
		}
		\label{f:SETUP}
	\end{figure}
	First we consider the dynamics starting from an artificial initial state
	to clearly see how large-scale vortex bundles generate vortices at a small scale.
	Four large-scale vortex bundles are imprinted in a uniform system,
	as shown in Fig.~\ref{f:SETUP} (a)~\cite{T.Yasuda}.
	Each bundle consists of six quantized vortex lines, expressed as
	\begin{eqnarray}
		\label{eq:SETUPEQ1}
		\tilde{\psi}({\bm r}) =
		\tilde{\psi}_0({\bm r}) \prod_{n=0}^5
			\frac{x - x_n \pm i(y - y_n)}
			{|x - x_n \pm i(y - y_n)|},
	\end{eqnarray}
	\begin{eqnarray}
		\label{eq:SETUPEQ2}
		\left(
			\begin{array}{l}
				x_n
				\\
				\\
				y_n
			\end{array}
		\right) =
		\left(
			\begin{array}{l}
				r_b \cos \left( \frac{n \pi}{3} \pm \frac{\pi z}{L} \right)
				\\
				\\
				r_b \sin \left( \frac{n \pi}{3} \pm \frac{\pi z}{L} \right)
			\end{array}
		\right), \ (n=0, 1, \cdots, 5),
	\end{eqnarray}
	where $\tilde{\psi}_0$ is a uniform wave function,
	$r_b$ is the bundle radius,
	and the rotation directions of the vortices are
	determined by the $\pm$ sign in Eq.~(\ref{eq:SETUPEQ1}).
	The time evolution of the system is obtained by solving Eq.~(\ref{eq:GPE})
	with $U=0$.
	To extract the vortex cores in Fig.~\ref{f:SETUP},
	we calculate the phase winding around each numerical mesh.
	In the time evolution, the two pairs of vortex bundles
	with opposite circulations first travel in the $\pm y$ directions.
	Because of the twisted configuration, fast spreading of
	the vortices in the bundles is avoided.
	At a later time, small vortex rings are created
	between the bundles, which are stretched to become ladder-like vortices
	perpendicular to the bundles, followed by the creation
	of new vortex rings, as shown in Fig.~\ref{f:SETUP}(b).

\begin{figure}[t]
		\centering
		\includegraphics[width=14cm]{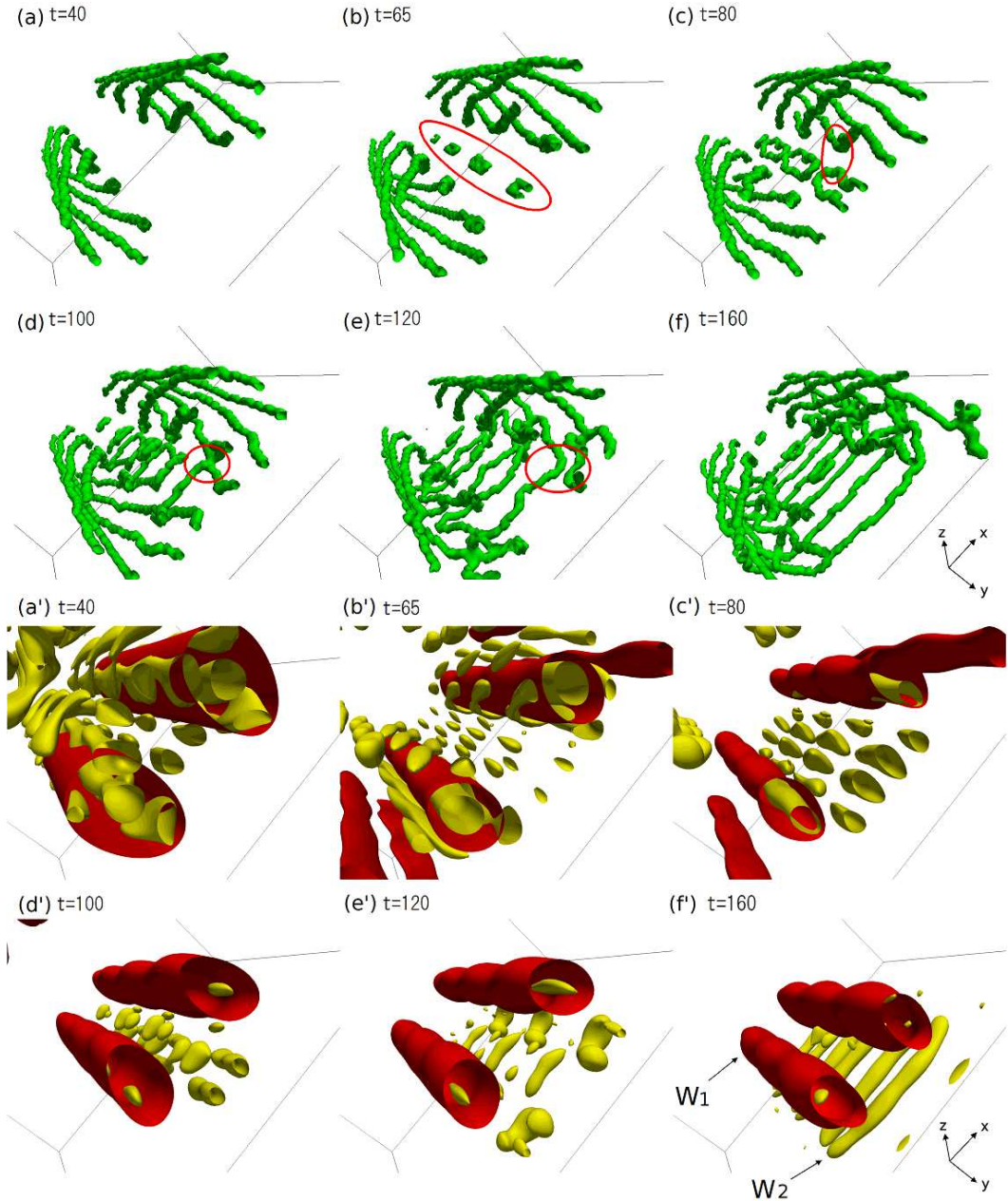}
		\caption{
			The same dynamics as in Fig.~\ref{f:SETUP}.
			The vortex cores are shown in (a)--(f),
			and the isodensity surfaces of band-pass filtered vorticity distributions
			$|\bm{W}_1|$ and $|\bm{W}_2|$ are shown in (a$'$)-(f$'$).
			The ranges of the wavenumbers in the Fourier filters are
			$4/\sqrt{2}\leq |\bm{k}| <4\sqrt{2}$ for $\bm{W}_1$ (red or dark gray)
			and $7/\sqrt{2}\leq |\bm{k}| < 7\sqrt{2}$ for $\bm{W}_2$ (yellow or light gray).
			The outline in (b) indicates the created vortex rings.
			The outlines in (c)--(e) highlight vortex reconnection.
		}
		\label{f:MODEL}
	\end{figure}

	Figure~\ref{f:MODEL} shows this process in detail.
	Small vortex rings are nucleated between the vortex bundles,
	as shown in Fig.~\ref{f:MODEL}(b),
	since the flow velocity in this region exceeds a local critical velocity.
    	The created vortex rings are then stretched and touch one
	of the vortices in the bundle, at which vortex reconnection occurs,
	as shown in Figs.~\ref{f:MODEL}(c)-~\ref{f:MODEL}(e).
	After the reconnection, the vortices form bridges
	between the bundles and a ladder structure is formed.
	Subsequently, new vortex rings are then created between the bundles,
	as shown in Fig.~\ref{f:MODEL}(f).
	This visualization of the vortex core dynamics
	is peculiar to quantum fluids,
	and clarifies the elementary process of vortex stretching.
	The band-pass filtered vorticity distributions
	$|\bm{W}_1|$ and $|\bm{W}_2|$ are shown
	in Figs.~\ref{f:MODEL}(a$'$)--~\ref{f:MODEL}(f$'$).
	The wavenumber ranges of the band-pass filters are
	$4/\sqrt{2}\leq |\bm{k}| <4\sqrt{2}$ for $\bm{W}_1$
	and $7/\sqrt{2}\leq |\bm{k}| < 7\sqrt{2}$ for $\bm{W}_2$;
	i.e., $\bm{W}_1$ and $\bm{W}_2$ are larger- and smaller-scale vorticity
	distributions, respectively.
	At $t=40$, the vorticities $\bm{W}_2$ are distributed in and around $\bm{W}_1$,
	as shown in Fig.~\ref{f:MODEL}(a$'$).
	As the vortices in the bundles expand, the distribution $\bm{W}_1$ diffuses,
	and the tube-like isodensity surfaces of $\bm{W}_1$ become thinner.
	Although the distribution of $\bm{W}_2$ is fragmented for $t \lesssim 120$,
	the vortex tubes of $\bm{W}_2$ that are orthogonal to those of $\bm{W}_1$ are
	established at $t=160$, as shown in Fig.~\ref{f:MODEL}(f$'$).
	These dynamics clearly show that a large-scale structure produces
	a small-scale structure, which causes the energy cascade.
	Similar dynamics are also observed in classical fluids~\cite{S.Goto2, M.V.Melander},
	which are attributed to vortex stretching.
	On the other hand, in the present case,
	the orthogonal structure is generated through
	stretching of quantized vortex rings and their reconnection~\cite{R.M.Kerr2}.
	\begin{figure}[t]
		\centering
		\includegraphics[width=16cm]{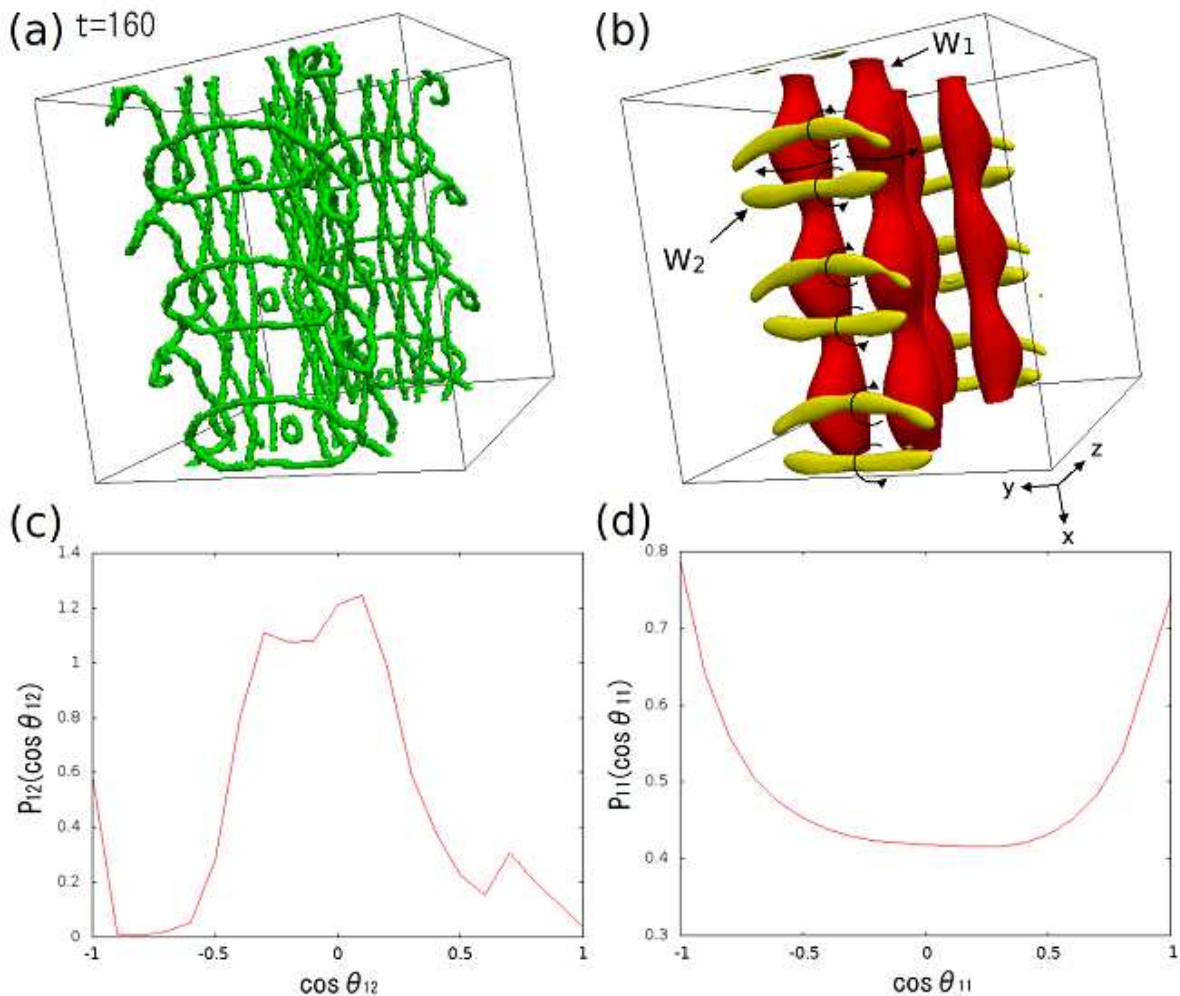}
		\caption{
			(a) Vortex cores., (b) Isodensity surfaces of $|\bm{W}_1|$ and $|\bm{W}_2|$.
			Normalized distribution (c) of angles $\theta_{12}$ between $\bm{W}_1(\bm{r})$ and
			$\bm{W}_2(\bm{r}+\Delta\bm{r})$,
			and (d) of angles $\theta_{11}$ between
			$\bm{W}_1(\bm{r})$ and $\bm{W}_1(\bm{r}+\Delta\bm{r})$ at $t=160$
			in the dynamics shown in Fig.~\ref{f:MODEL}.
			The curly arrows in (b) represent the directions of the circulations.
			In (c) and (d), the data are taken for the range $48 \leq |\Delta\bm{r}| < 50$.
		}
		\label{f:ORTHANT}
	\end{figure}
	Figures~\ref{f:ORTHANT}(a) and ~\ref{f:ORTHANT}(b) show the vortex-core distribution
	and band-pass filtered vortex distributions of the ladder structure in Figs.~\ref{f:MODEL}(f)
	and ~\ref{f:MODEL}(f$'$) seen from different angles.
	The rotation directions of ${\bm W}_1$ and ${\bm W}_2$ are shown in Fig.~\ref{f:ORTHANT}(b),
	indicateing that the small-scale vortex tubes have opposite rotation directions.
	To quantify the distributions of the angles between the vortex tubes,
	we calculate the distributions of $\cos \theta_{12}$ and $\cos \theta_{11}$ for
	the state in Fig.~\ref{f:ORTHANT}(b),
	which are shown in Figs.~\ref{f:ORTHANT}(c) and ~\ref{f:ORTHANT}(d).
	The distance $|\Delta\bm{r}|$ in Eqs.~(\ref{eq:ORTH}) and (\ref{eq:ANTI})
	is taken to be $48$--$50$, which is the typical distance
	for the distances between vortex tubes in Fig.~\ref{f:ORTHANT}(b).
	The distribution $P_{12}(\cos \theta_{12})$ is large around $\cos \theta_{12}=0$,
	which indicates that $\bm{W}_1(\bm{r})$ and $\bm{W}_2(\bm{r}+\Delta\bm{r})$
	tend to be orthogonal to each other.
	The distribution $P_{11}(\cos\theta_{11})$ is large at $\cos\theta_{11}=-1$,
	which indicates that $\bm{W}_1(\bm{r})$ and $\bm{W}_1(\bm{r}+\Delta\bm{r})$
	tend to be antiparallel with each other.
	There is also a peak at $\cos\theta_{11}=1$, due to case in which
	$\bm{W}_1(\bm{r})$ and $\bm{W}_1(\bm{r}+\Delta\bm{r})$ are in the same
	vortex tube.
	These results are similar to those in classical fluids with a similar setup
	~\cite{S.Goto2}.
	\subsection{Fully-developed isotropic turbulence}
	\label{s:distributions}
	\begin{figure}[t]
		\centering
		\includegraphics[width=17cm]{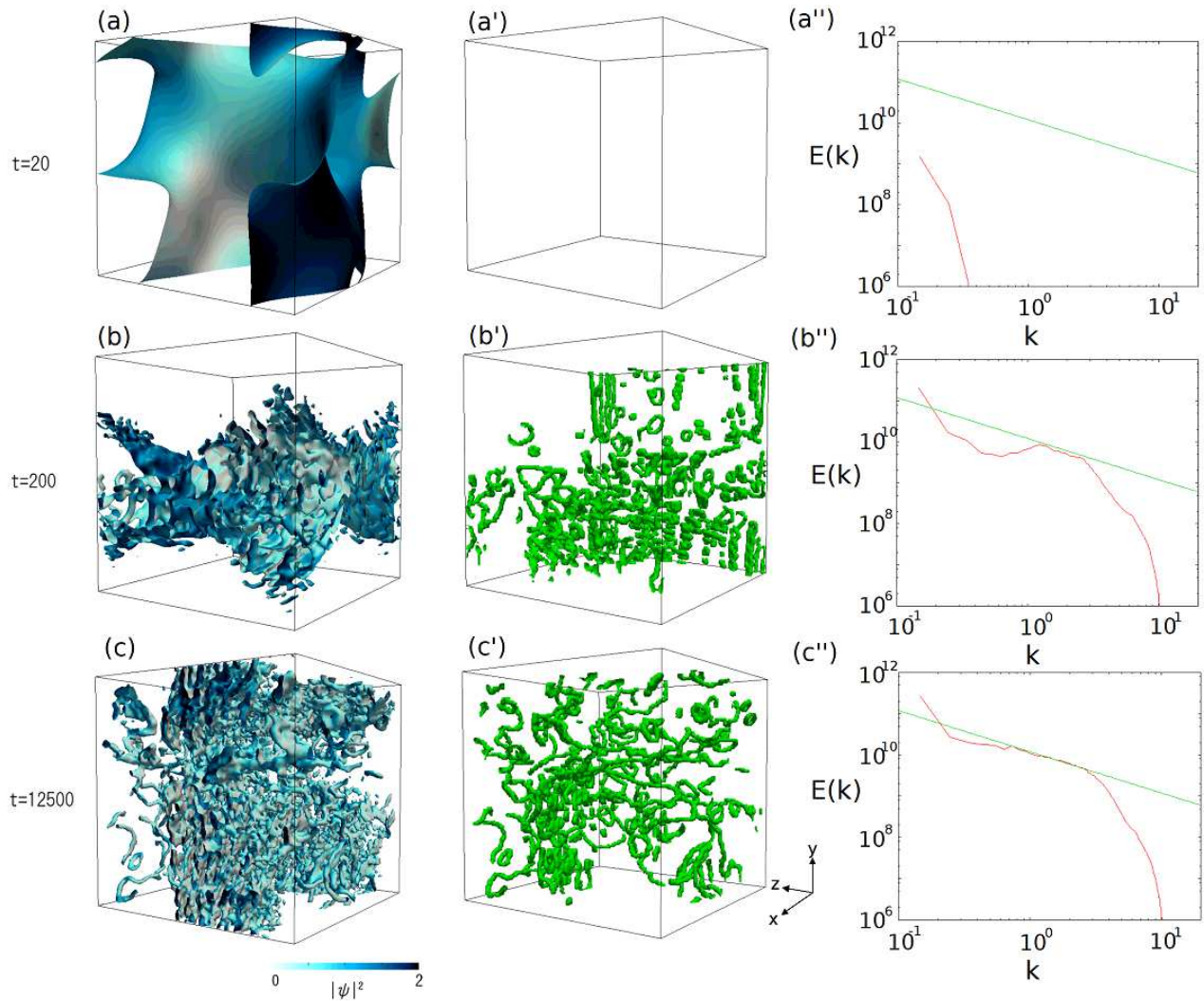}
		\caption{
			(a)--(c) Isodensity surfaces of $|\tilde{\psi}|^2$,
			(a$'$)--(c$'$) vortex-core profiles,
			and (a$''$)--(c$''$) power spectra $E(k)$ (arbitrary units) for the dynamics driven by
			the random potential $U(\bm{r}, t)$ in Eq.~(\ref{eq:POTENTIAL}) with
			$\kappa=0.05, A_0=0.5$, and $l=8\pi $.
			The size of the cubes in (a)--(c) and (a$'$)--(c$'$) is  $128^3$.
			The slope of the lines in (a$''$)--(c$''$) is $-5/3$.
		}
		\label{f:TURB}
	\end{figure}
	We consider here the case of isotropic quantum turbulence.
	We numerically solve Eq.~(\ref{eq:NGPE})
	with a time-dependent random potential $U(\bm{r},t)$ generated by
	the method given in Appendix~\ref{s:random}.
	The initial condition is the homogeneous state,
	and the system evolves until the steady turbulent state is reached.
	The isodensity surfaces of $|\tilde\psi|^2$,
	the vortex-core profiles,
	and the power spectra $E(k)$ are shown
	in Figs.~\ref{f:TURB}(a)--\ref{f:TURB}(c), ~\ref{f:TURB}(a$'$)--\ref{f:TURB}(c$'$),
	and ~\ref{f:TURB}(a$''$)--\ref{f:TURB}(c$''$), respectively.
	The power spectrum $E(k)$ is defined in Appendix~\ref{s:pspectrum}.
	Since the characteristic spatial scale of the random potential is
	of the order of the system size,
	long-wavelength modes are excited at $t=20$.
	An energy cascade from the long-wavelength to short-wavelength modes then occurs.
	Kolmogorov's power law,
	$E(k) \propto k^{-5/3}$, is observed in Fig.~\ref{f:TURB}(c$''$),
	and the system is in the turbulent state.
	\begin{figure}[t]
		\centering
		\includegraphics[width=13cm]{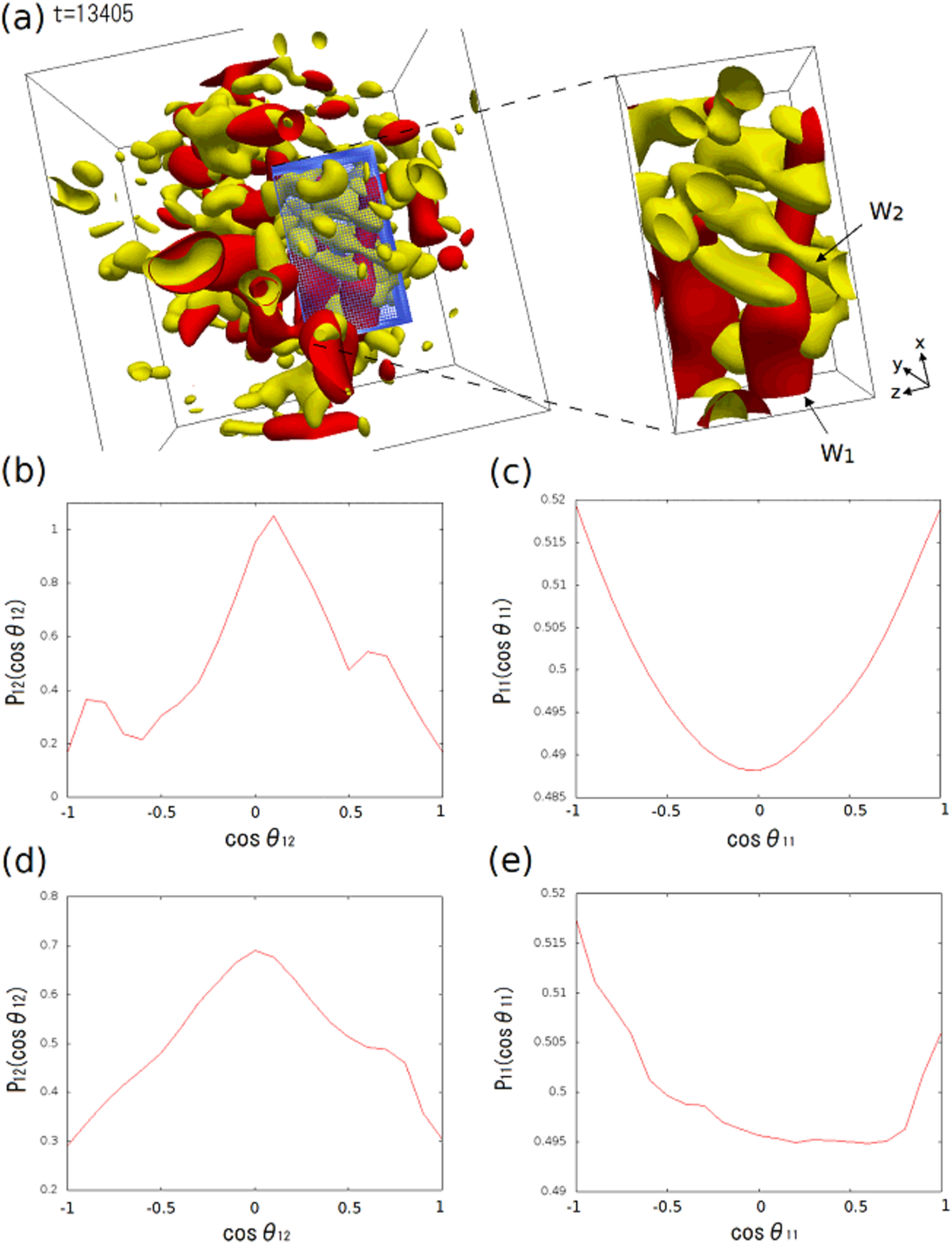}
		\caption{
			Vorticity distribution of fully-developed turbulence for the dynamics
			shown in Fig.~\ref{f:TURB}.
			(a) Isodensity surfaces of vorticity distributions $|\bm{W}_1|$
			and $|\bm{W}_2|$ at $t=13405$, where the wavenumber ranges of
			the Fourier filtering are the same as those in Fig.~\ref{f:MODEL}.
			The magnified region clearly shows that $\bm{W}_1$ and $\bm{W}_2$
			tend to be orthogonal to each other.
			(b) Distribution of angles $\theta_{12}$ between $\bm{W}_1(\bm{r})$
			and $\bm{W}_2(\bm{r}+\Delta\bm{r})$.
			(c) Distribution of angles $\theta_{11}$ between $\bm{W}_1(\bm{r})$
			and $\bm{W}_1(\bm{r}+\Delta\bm{r})$.
			(b) and (c) were obtained by a single shot at
			$t=13405$ and (d) and (e) are averages of $1000$ shots for
			$13000 \leq t < 14000$.
			The distributions $P_{11}$ and $P_{12}$ are calculated over the whole
			$128^3$ region (not restricted to the selected region as
			in the right panel in (a)).
		}
		\label{f:ANTIPARA}
	\end{figure}

	To investigate how the energy is transfered from large to small scales,
	we calculate the band-pass filtered vorticity distributions $\bm{W}_1$ and $\bm{W}_2$.
	Figure~\ref{f:ANTIPARA}(a) shows
	the isodensity surfaces of
	$|\bm{W}_1|$ and $|\bm{W}_2|$ at $t=13405$.
	The definitions of $\bm{W}_1$ and $\bm{W}_2$
	are the same as those in Sec.~\ref{s:nucleation}, i.e.,
	$\bm{W}_1$ and $\bm{W}_2$ correspond to larger- and smaller-scale vorticity
	distributions, respectively.
	The right-hand panel in Fig.~\ref{f:ANTIPARA}(a) shows
	an enlarged view of the meshed region.
	In the enlarged view, we can clearly see that
	the pair of vortex tubes in $\bm{W}_1$ aligns in parallel
	and the vortex tubes in $\bm{W}_2$ tend to be orthogonal to those in $\bm{W}_1$.
	This configuration of vortex tubes in $\bm{W}_1$ and $\bm{W}_2$ is
	similar to that in Fig.~\ref{f:ORTHANT}(b) in which the vortex bundles are artificially generated.
	By contrast, we note that the structures shown in Fig.~\ref{f:ANTIPARA}(a) are formed by a random potential.
	Figures~\ref{f:ANTIPARA}(b) and ~\ref{f:ANTIPARA}(c) show
	the angular distributions $P_{12}$ and $P_{11}$
	of the vorticities $\bm{W}_1$ and $\bm{W}_2$,
	defined in Eqs.~(\ref{eq:ORTH}) and (\ref{eq:ANTI}).
	These were calculated for the whole $L^3=128^3$ region.
	Although the turbulent state is induced by a random potential,
	there are significant correlations between the vorticities.
	The vorticities $\bm{W}_1(\bm{r})$ and
	$\bm{W}_2(\bm{r} + \Delta \bm{r})$ tend to be orthogonal to each other,
	and $P_{12}(\cos \theta_{12})$ has a peak at $\cos \theta_{12} = 0$.
	The vorticities $\bm{W}_1$ at $\bm{r}$ and $\bm{r} + \Delta \bm{r}$
	tend to be antiparallel with each other, and $P_{11}(\cos \theta_{11})$
	has a peak at $\cos \theta_{11} = -1$.
	(The peak at $\cos\theta_{11}=1$ is due to the correlation within
	a single vortex tube.)
	To assure that these tendencies are not incidental,
	we calculate the time-averaged angular distributions for $13000 \leq t < 14000$.
	The characteristic time scale of the random potential is $\kappa^{-1} = 20$,
	which is long enough to observe the ensemble averaged behaviors.
	We find that the tendencies in Figs.~\ref{f:ANTIPARA}(d) and
	~\ref{f:ANTIPARA}(e) are the same as those in Figs.~\ref{f:ANTIPARA}(b)
	and ~\ref{f:ANTIPARA}(c), respectively, and therefore the above angular correlations
	in the vorticity distributions can be observed constantly.

	Thus, we have shown that in quantum turbulence, large-scale vorticity
	$\bm{W}_1$ tends to have antiparallel structures
	and small-scale vorticity $\bm{W}_2$ tends to be perpendicular to $\bm{W}_1$.
	These results indicate that the energy is transferred from large to small scales
	through vortex stretch dynamics,
	which implies that this is one of the mechanisms of the energy cascade
	and emergence of Kolmogorov's law in quantum turbulence.

	\section{CONCLUSIONS}
	\label{s:conclusions}
	We have investigated the dynamics of vortices in quantum fluids
	using the numerical simulation of the Gross-Pitaevskii equation.
	We defined band-pass filtered vorticity distributions to study the
	dynamics at each scale.
	In Sec.~\ref{s:nucleation},
	we examined the dynamics of the vortex bundles
	and observed that large-scale antiparallel vortices nucleate
	small-scale vortices orthogonal to those at the large scale.
	These processes are induced by nucleation of quantized vortex rings
	and their reconnections.
	In Sec.~\ref{s:distributions},
	we applied our method to
	the homogeneous isotropic turbulent state.
	Despite the fact that the turbulent state is generated by a random potential,
	there are significant correlations in the vorticity distributions.
	We found that intra-scale vorticities tend to align in antiparallel
	and the smaller-scale vortices tend to be orthogonal to larger-scale vortices.
	These vortex dynamics may play an important role
	in the energy cascade and Kolmogorov's law in quantum turbulence.

	In the present study, we have only considered vorticity distributions
	at two scales $\bm{W}_1$ and $\bm{W}_2$.
	Performing numerical simulations in a larger system will provide
	vorticity distributions at multiple scales, which will reveal the multistage
	generation of antiparallel and orthogonal vortices.

	\begin{acknowledgments}
		The present study was supported by
		JSPS KAKENHI Grant Numbers JP16K05505, JP17K05595, and JP17K05596.
	\end{acknowledgments}

	\appendix
	\section{Time-dependent random potential to generate quantum turbulence}
	\label{s:random}
	To generate homogeneous isotropic quantum turbulence,
	we use a random potential.
	The potential is expanded as
	\begin{eqnarray}
		\label{eq:POTENTIAL}
		U( {\bm r}, t) = \sum_{\bm k} C_{{\bm k}} (t) e^{i {\bm k} \cdot {\bm r}}.
	\end{eqnarray}
	The time-dependent Fourier components $C_{\rm{k}}(t)$ follow
	the Langevin equation
	\begin{eqnarray}
		\label{eq:LANGEVIN}
		\frac{{\rm d} C_{\bm k}( t)}{{\rm d}t} = -\kappa C_{\bm k}(t) + f_{\bm k}(t),
	\end{eqnarray}
	where the constant $\kappa > 0$ determines the time scale
	of potential variation
	and $f_{\bm k}(t)$ is the Gaussian noise with an average,
	\begin{eqnarray}
		\left < f_{\bm k}(t) \right> = 0,
	\end{eqnarray}
	and correlation function,
	\begin{eqnarray}
		\left < f_{\bm k}(t) f_{{\bm k}^\prime}(t^\prime)\right> =
		A_{\bm k}\delta_{{\bm k}{\bm k}^\prime}\delta(t-t^\prime).
	\end{eqnarray}
	The magnitude $A_{\bm k}$ is given by
	\begin{eqnarray}
		\label{eq:RANDOM}
		A_{\bm k} = A_0 e^{-\left(\frac{l}{2}|\bm{k}|\right)^2},
	\end{eqnarray}
	where the parameter $l$ determines the characteristic scale
	of the random potential.
	Using the solution of the Langevin equation in Eq.~(\ref{eq:POTENTIAL}), we have
	\begin{eqnarray}
		\left<
			C_{\bm k}(t)
			C_{{\bm k}^\prime}(t^\prime)
		\right>
		= \frac{A_{\bm k}}{2 \kappa} \delta_{{\bm k}{\bm k}^\prime} e^{-\kappa|t-t^\prime|},
	\end{eqnarray}
	which gives
	\begin{eqnarray}
		\left<
			U({\bm r}, t) U({\bm r}^\prime, t^\prime)
		\right> \propto
		e^{-\kappa |t-t'|} e^{|\bm{r}-\bm{r}'|^2/l^2}.
	\end{eqnarray}
	In the numerical simulation in Sec.~\ref{s:distributions},
	the coefficients $C_{\bm k}(t)$
	numerically evolve according to Eq.~(\ref{eq:LANGEVIN}). The inverse Fourier transform
	in Eq.~(\ref{eq:POTENTIAL}) thus gives the time-dependent random potential
	with spatial and temporal scales of $l$ and $\kappa^{-1}$, respectively.

	\section{Incompressible kinetic-energy power spectrum}
	\label{s:pspectrum}
	The kinetic energy of a quantum fluid is expressed as
	\begin{eqnarray}
		E_{{\rm kinetic}} & = & -\frac{1}{2} \int {\rm d}\bm{r} \tilde{\psi}^* \tilde{\nabla}^2 \tilde{\psi},
		\\
		& = & \frac{1}{2}\int {\rm d}\bm{r}|\tilde{\nabla} \tilde{\psi}|^2.
	\end{eqnarray}
	Using the transformation, $\tilde{\psi}(\bm{r}) = \sqrt{\rho(\bm{r})} e^{i \phi(\bm{r})}$,
	the kinetic-energy can be divided into two terms as
	\begin{eqnarray}
		E_{{\rm kinetic}} & = & \frac{1}{2}\int {\rm d}\bm{r}\left[
		\rho(\tilde{\nabla} \phi)^2 + (\tilde{\nabla} \sqrt{\rho})^2
		\right],
		\\
		& = & E_1 + E_2,
	\end{eqnarray}
	where $E_1=\frac{1}{2}\int {\rm d}\bm{r} \rho(\tilde{\nabla} \phi)^2$
	corresponds to the classical kinetic energy
	and $E_2=\frac{1}{2}\int {\rm d}\bm{r} (\tilde{\nabla} \sqrt{\rho})^2$ comes from the quantum pressure.
	We define $\bm{w}(\bm{r})=\sqrt{\rho(\bm{r})}\tilde{\nabla}\tilde{\phi}(\bm{r})$
	and its Fourier transform,
	\begin{eqnarray}
		\tilde{\bm{w}}(\bm{k}) = \int \bm{w}(\bm{r}) e^{-i \bm{k} \cdot \bm{r}} {\rm d} \bm{r}.
	\end{eqnarray}
	The field $\bm{w}$ is divided into compressible and incompressible parts as
	\begin{eqnarray}
		\tilde{\bm{w}}(\bm{k}) & = &
		\frac{\bm{k}\cdot\tilde{\bm{w}}(\bm{k})}{k^2}\bm{k}
		+ \frac{(\bm{k}\times\tilde{\bm{w}})\times\bm{k}}{k^2},
		\\
		\tilde{\bm{w}}_{\rm{L}} & = &
		\frac{\bm{k}\cdot\tilde{\bm{w}}(\bm{k})}{k^2}\bm{k},
		\\
		\tilde{\bm{w}}_{\rm{T}} & = &
		\frac{(\bm{k}\times\tilde{\bm{w}})\times\bm{k}}{k^2}.
	\end{eqnarray}
	The kinetic energy can be rewritten as
	\begin{eqnarray}
		E_1 & = & \frac{1}{2} \int {\rm d}\bm{r}\left|\bm{w}(\bm{r})\right|^2,
		\\
		& = & \frac{1}{2}\int {\rm d}\bm{r} \left[
			\left| \bm{w}_{{\rm T}}(\bm{r})\right|^2
			+ \left| \bm{w}_{{\rm L}}(\bm{r})\right|^2
		\right].
	\end{eqnarray}
	We focus on the incompressible part of the kinetic energy,
	\begin{eqnarray}
		E_1^{{\rm ic}}
		& = & \frac{1}{2} \int \left|\bm{w}_{{\rm T}}(\bm{r}) \right|^2 {\rm d}\bm{r},
		\\
		& = & \frac{1}{2}\int\frac{{\rm d}{\bm k}}{(2 \pi)^3}
		\tilde{\bm{w}}_{{\rm T}}(\bm{k}) \cdot \tilde{\bm{w}}_{{\rm T}}(-\bm{k}),
		\\
		& = & \int {\rm d}k E(k),
	\end{eqnarray}
	which is the definition of the power spectrum $E(k)$ of the incompressible flow.

	

\begin{thebibliography}{99}
		\bibitem{O.Reynolds}
		O. Reynolds,
		An Experimental Investigation of the Circumstances Which Determine
		Whether the Motion of Water Shall Be Direct or Sinuous
		and of the Law of Resistance in Parallel Channels,
		Phil. Trans. R. Soc. London \textbf{174}, 935 (1883).

		\bibitem{L.F.Richardson}
		L. F. Richardson,
		{\it Weather Prediction by Numerical Process},
		(Cambridge University Press, Cambridge, 1922).

		\bibitem{U.Frisch1}
		U. Frisch, {\it Turbulence: The Legacy of A. N. Kolmogorov},
		(Cambridge University Press, Cambridge, 1995).

		\bibitem{A.N.Kolmogorov}
		A. N. Kolmogorov,
		The local structure of turbulence
		in incompressible viscous fluid for very large Reynolds numbers,
		Dokl. Akad. Nauk SSSR \textbf{30}, 301 (1941).

		\bibitem{G.K.Batchelor}
		G. K. Batchelor, and I. Proudman,
		The effect of rapid distortion of a fluid in turbulent motion,
		Quart. J. Mech. Appl. Math. \textbf{7}, 83 (1954).

		\bibitem{T.Tatsumi}
		T. Tatsumi,
		The Theory of Decay Process of Incompressible Isotropic Trubulence,
		Proc. R. Soc. London \textbf{239}, 16 (1957).

    	\bibitem{R.H.Kraichinan}
    	R. H. Kraichinan,
    	On Kolmogorov's inertial-range theories,
    	J. Fluid Mech. \textbf{62}, 305 (1974).

		\bibitem{U.Frisch2}
		U. Flisch, P. L. Sulem, and M. Nelkin,
		A simple dynamical model of intermittent fully developed turbulence,
		J. Fluid Mech. \textbf{87}, 719 (1978).

        \bibitem{S.Goto1}
        S. Goto,
        A physical mechanism of the energy cascade
		in homogeneous isotropic turbulence,
        J. Fluid Mech. \textbf{605}, 355 (2008).

		\bibitem{S.Goto2}
		S. Goto, Y. Saito, and G. Kawahara,
		Hierarchy of antiparallel vortex tubes
		in spatially periodic turbulence at high Reynolds numbers,
		Phys. Rev. Fluids \textbf{2}, 064603 (2017).

		\bibitem{K.Sasaki1}
		K. Sasaki, N. Suzuki, and H. Saito,
		B\'enard-von K\'arm\'an Vortex Street in a Bose-Einstein Condensate,
		Phys. Rev. Lett. \textbf{104}, 150404 (2010).

		\bibitem{M.T.Reeves1}
		M. T. Reeves, T. P. Billam, B. P. Anderson, and A. S. Bradley,
		Identifying a Superfluid Reynolds Number via Dynamical Similarity,
		Phys. Rev. Lett. \textbf{114}, 155302 (2015)

		\bibitem{W.J.Kwon}
		W. J. Kwon, J. H. Kim, S. W. Seo, and Y. Shin,
		Observation of von K\'arm\'an Vortex Street in an Atomic Superfluid Gas,
		Phys. Rev. Lett. \textbf{117}, 245301 (2016).

		\bibitem{K.Sasaki2}
		K. Sasaki, N. Suzuki, D. Akamatsu, and H. Saito,
		Rayleigh-Taylor instability and mashroom-pattren formation
		in a two-component Bose-Einstein condensate,
		Phys. Rev. A \textbf{80}, 063611 (2009).

		\bibitem{T.Kadokura}
		T. Kadokura, T. Aioi, K. Sasaki, T. Kishimoto, and H. Saito,
		Rayleigh-Talor instability in a two-component Bose-Einstein
		condensate with rotational symmetry,
		Phys. Rev. A \textbf{85}, 013602 (2012).

		\bibitem{H.Takeuchi}
		H. Takeuchi, N. Suzuki, K. Kasamatsu, H. Saito, and M. Tsubota,
		Quantum Kelvin-Helmholtz instability in phase-separated
		two-component Bose-Einstein condensates,
		Phys. Rev. B \textbf{81}, 094517 (2010).

		\bibitem{A.Bezett}
		A. Bezett, V. Bychkov, E. Lundh, D. Kobyakov, and M. Marklund,
		Magnetic Richtmyer-Meshkov instability in a two-component
		Bose-Einstein condensate,
		Phys. Rev. A \textbf{82}, 043608 (2010).

		\bibitem{L.Skrbek}
		L. Skrbek,
		Quantum turbulence,
		J. Phys. Conf. \textbf{318}, 012004 (2011).

		\bibitem{N.Navon}
		N. Navon, A. I. Gaunt, R. P. Smith, and Z. Hadzibabic,
		Emergence of a turbulent cascade in a quantum gas,
		Nature (London) \textbf{539}, 3 (2016).

		\bibitem{M.Tsubota1}
		T. Araki, M. Tsubota, and S. K. Nemirovskii,
		Energy Spectrum of Superfluid Turbulence with No Normal-Fluid Component,
		Phys. Rev. Lett. \textbf{89}, 145301 (2002).

		\bibitem{M.Kobayashi1}
		M. Kobayashi, and M. Tsubota,
		Kolmogorov Spectrum of Superfluid Turbulence:
		Numerical Analysis of the Gross-Pitaevskii Equation
		with a Small-Scale Dissipation,
		Phys. Rev. Lett. \textbf{94}, 065302 (2005).

		\bibitem{M.Kobayashi2}
		M. Kobayashi, and M. Tsubota,
		Thermal Dissipation in Quantum Turbulence,
		Phys. Rev. Lett. \textbf{97}, 145301 (2006).

		\bibitem{M.Kobayashi3}
		M. Kobayashi, and M. Tsubota,
		Quantum turbulence in a trapped Bose-Einstein condensate,
		Phys. Rev. A \textbf{76}, 045603 (2007).

		\bibitem{A.W.Baggaley}
		A. W. Baggaley, J. Laurie, and C. F. Barenghi,
		Vortex-Density Fluctuations, Energy Spectra,
		and Vortical Regions in Superfluid Turbulence,
		Phys. Rev. Lett. \textbf{109}, 205304 (2012).

		\bibitem{M.Tsubota2}
		M. Tsubota,
		Quantum turbulence: from superfluid helium to atomic Bose-Einstein condensates,
		Contemp. Phys. \textbf{50}, 463 (2009).

		\bibitem{K.Fujimoto}
		M. Tsubota, K. Fujimoto, and S. Yui,
		Numerical Studies of Quantum Turbulence,
		J. Low Temp. Phys. \textbf{188}, 119 (2017).

		\bibitem{P.J.Tabeling}
		J. Paret and P. Tabeling,
		Experimental Observation of the Two-Dimensional Inverse Energy Cascade,
		Phys. Rev. Lett. \textbf{79}, 4162 (1997).

		\bibitem{A.C.White}
		A. C. White, B. P. Anderson, and V. S. Bagnato,
		Vortices and turbulence in trapped atomic condensates,
		Proc. Natl. Acad. Sci. U.S.A. \textbf{111}, 4719 (2014).

		\bibitem{A.Villois}
		A. Villois, D. Proment, and G. Krstulovic,
		Evolution of a superfluid vortex filament tangle driven
		by the Gross-Pitaevskii equation,
		Phys. Rev. E \textbf{93}, 061103(R) (2016).

		\bibitem{P.M.Walmsley}
		P. M. Walmsley, A. I. Golov, H. E. Hall, A. A. Levchenko, and W. F. Vinen,
		Dissipation of Quantum Turbulence in the Zero Temperature Limit,
		Phys. Rev. Lett. \textbf{99}, 265302 (2007).

		\bibitem{M.T.Reeves2}
		M. T. Reeves, B. P. Anderson, and A. S. Bradley,
		Classical and quantum regimes of two-dimensional turbulence
		in trapped Bose-Einstein condensates,
		Phys. Rev. A. \textbf{86}, 053621 (2012).

		\bibitem{C.F.Barenghi1}
		D. Kivotides, J. C. Vassilicos, D. C. Samuels, and C. F. Barenghi,
		Kelvin Waves Cascade in Superfluid Turbulence,
		Phys. Rev. Lett. \textbf{86}, 3080 (2001).

		\bibitem{C.F.Barenghi2}
		C. F. Barenghi, V. S. L'vov, and P. E. Roche,
		Experimental, numerical, and analytical velocity spectra in turbulent quantum fluid,
		Proc. Natl. Acad. Sci. U.S.A. \textbf{111}, 4683 (2014).

		\bibitem{M.Kursa}
		M. Kursa, K. Bajer, and T. Lipniacki,
		Cascade of vortex loops initiated by a single reconnection of quantum vortices,
		Phys. Rev. B \textbf{83}, 014515 (2011).

        \bibitem{R.M.Kerr1}
        R. M. Kerr,
        Swirling, turbulent vortex rings formed from a chain reaction of reconnection events,
        Phys. Fluids \textbf{25}, 065101 (2013).

		\bibitem{S.R.Stalp}
		S. R. Stalp, L. Skrbek, and R. J. Donnelly,
		Decay of Grid Turbulence in a Finite Channel,
		Phys. Rev. Lett. \textbf{82}, 4831 (1999).

		\bibitem{K.W.Madison}
		K. W. Madison, F. Chevy, W. Wohlleben, and J. Dalibard,
		Vortex Formation in a Stirred Bose-Einstein condensate,
		Phys. Rev. Lett. \textbf{84}, 806 (2000).

		\bibitem{N.G.Parker}
		N. G. Parker, and C. S. Adams,
		Emergence and Decay of Turbulence in Stirred Atomic Bose-Einstein condensates,
		Phys. Rev. Lett. \textbf{95}, 145301 (2005).

		\bibitem{C.Nore}
		C. Nore, M. Abid, and M. E. Brachet,
		Kolmogorov Turbulence in Low-Temperature Superflows,
		Phys. Rev. Lett. \textbf{78}, 3896 (1997).

		\bibitem{M.Kobayashi4}
		M.Tsubota, and M.Kobayashi,
		Quantum Turbulence in Trapped Atomic Bose-Einstein condensates,
		J. Low Temp. Phys. \textbf{150}, 402 (2008).

		\bibitem{S.Goto3}
		S. Goto,
		Coherent Structures and Energy Cascade in Homogeneous Turbulence,
		Prog. Theor. Phys. Suppl. \textbf{195}, 139 (2012).

		\bibitem{T.Yasuda}
		T. Yasuda, S. Goto, and G. Kawahara,
		Quasi-cyclic evolution of turbulence driven by a steady force in a periodic cube,
		Fluid Dyn. Res. \textbf{46}, 061413 (2014).

		\bibitem{M.V.Melander}
		M. V. Melander, and F. Hussain,
		Core dynamics on a vortex column,
		Fluid Dyn. Res. \textbf{13}, 1 (1994).

		\bibitem{R.M.Kerr2}
		R. M. Kerr,
		Vortex Stretching as a Mechanism of Quantum Kinetic Energy Decay,
		Phys. Rev. Lett. \textbf{106}, 224501 (2011).



	\end{thebibliography}
\end{document}